\documentclass[12pt,a4paper,aps,prd,preprint,superscriptaddress,nofootinbib]{revtex4-1}
\usepackage[utf8]{inputenc}
\usepackage{graphicx}
\usepackage{amssymb}
\usepackage{textcomp}
\usepackage{amsmath}
\usepackage{tabularx}
\usepackage{bm}
\usepackage{times}
\usepackage{color}
\usepackage{slashed}
\usepackage{multirow}
\usepackage{verbatim}
\usepackage{cancel}
\usepackage{subfigure}
\usepackage[normalem]{ulem}

\usepackage[colorlinks=true, pdfstartview=FitV, linkcolor=blue, citecolor=blue, urlcolor=blue]{hyperref}
\allowdisplaybreaks[4]

\def\lsim{\mathrel{\raise.3ex\hbox{$<$\kern-.75em\lower1ex\hbox{$\sim$}}}}
\def\gsim{\mathrel{\raise.3ex\hbox{$>$\kern-.75em\lower1ex\hbox{$\sim$}}}}

\newcommand{\bew}{\begin{widetext}}
\newcommand{\enw}{\end{widetext}}
\newcommand{\bee}{\begin{equation}}
\newcommand{\ene}{\end{equation}}
\newcommand{\bea}{\begin{eqnarray}}
\newcommand{\ena}{\end{eqnarray}}
\newcommand{\bes}{\begin{subequations}}
\newcommand{\ens}{\end{subequations}}

\def\calm{\mathcal{M}}

\definecolor{orange}{rgb}{1,0.5,0}

\begin{document}

\title{Exploring millicharged particles in laboratory and astrophysical strong-field regimes}

\author{Cheng-Rui Jiang}
\email{jiangcr@mail.nankai.edu.cn}
\affiliation{
School of Physics, Nankai University, Tianjin 300071, China
}

\author{Tong Li}
\email{litong@nankai.edu.cn}
\affiliation{
School of Physics, Nankai University, Tianjin 300071, China
}

\author{Kai Ma}
\email{kai@xauat.edu.cn}
\affiliation{Faculty of Science, Xi'an University of Architecture and Technology, Xi'an, 710055, China}

\author{Haolong Wang}
\email{2120250183@mail.nankai.edu.cn}
\affiliation{
School of Physics, Nankai University, Tianjin 300071, China
}

\begin{abstract}
The probe of light dark particles beyond the Standard Model (SM) under a strong-field environment has drawn significant attention. In this work, we investigate the potential to search for and constrain light millicharged particles (MCPs) via strong
electromagnetic fields in both laboratory laser experiments and astrophysical environments such
as magnetars. We propose the MCP pair production from
nonlinear Compton scattering through the interaction of a relativistic electron beam with a high-intensity laser pulse. The Furry picture
and Volkov solution of Dirac equation in a background electromagnetic field are used to describe the electrons and MCPs under an external classical laser field. We calculate the cross sections of nonlinear
Compton scattering to MCP pairs and take into account the  irreducible SM background with missing neutrinos. We also revisit the MCP pair production via the Schwinger mechanism from magnetars with ultra-strong magnetic field and parallel electric field in polar gap. The energy loss due to the Schwinger pair production of MCPs and electric field acceleration is evaluated based on Ruderman-Sutherland model for confirmed magnetars. We find that the constraints from highly magnetized magnetars and the search
potential in laboratory laser experiments are complementary.
\end{abstract}

\maketitle
\tableofcontents

\section{Introduction}
\label{sec:Intro}

Ultra-strong field has been catalyzing novel exploration for particle physics and astrophysics.
In 1951, J.~Schwinger pointed out that at an external field strength of $\mathcal{E}\simeq 1.32\times 10^{18}~{\rm V/m}$, the quantum electrodynamics (QED) vacuum becomes unstable and the virtual electron-positron pair
fluctuations therein can be converted into real electron-positron pairs~\cite{Schwinger:1951nm}.
The extreme electromagnetic environments with field strengths approaching the Schwinger critical threshold offer a unique frontier for testing QED beyond the perturbative regime and probing other exotic physics.

Standard QED calculations rely on the small fine-structure constant $\alpha\approx 1/137$ to yield convergent perturbative expansions. In contrast, strong fields induce intrinsically non-perturbative and nonlinear effects in such as electron-positron pair production. This is first theoretically proposed for laser-driven systems~\cite{BunkinTugov1970}, and now routinely invoked to explain pair cascades in pulsar~\cite{Daugherty:1976mg,Hibschman:2001wk,Harding:2006qn} and magnetar~\cite{Thompson:2008pp,Medin:2010tp,Thompson:2020hwt,Harding:2025vje} magnetospheres. Strong field also enables precision tests of the Standard Model (SM) and searches for new physics beyond the SM. For instance, vacuum birefringence is driven by modifications to photon propagation caused by strong-field QED vacuum polarization~\cite{Brezin:1971nd,Lai:2001di,Wang:2009sg,Kim:2024npq}. Laboratory laser experiments and astrophysical magnetar emission are also ideal for detecting dark sector candidates, e.g., the production of dark particles from laser-assisted nonlinear scattering processes~\cite{Fuchs:2024edo,Dillon:2018ypt,King:2018qbq,Bai:2021gbm,Dillon:2018ouq,King:2019cpj,Beyer:2021mzq,Huang:2020lxo,Ma:2024ywm,Ma:2025axq,Ma:2025ymt,Li:2025ptt} or enhanced conversion signals via axion-photon oscillation in magnetar magnetospheres~\cite{Hook:2018iia,Huang:2018lxq,Long:2024qvd}.

Recently, the search for light dark particles beyond the SM has stimulated considerable interest across multiple disciplines. Millicharged particles (MCPs, denoted as $\chi$ below) as one class of such particles have been the focus of research. These hypothetical fermions have an electric charge far smaller than that of an electron $e$. Their charge is typically parameterized as $q_\chi=\epsilon e$, where $\epsilon$ is a dimensionless charge fraction much less than 1. The interaction Lagrangian between the MCP and the SM photon field $A_\mu$ becomes
\begin{eqnarray}
\epsilon e A_\mu \overline{\chi} \gamma^\mu \chi\;.
\end{eqnarray}
These particles arise naturally in theoretical extensions of the SM. A common framework involves a ``hidden sector'' containing new particles that interact via a new $U(1)$ gauge force (a ``dark photon'')~\cite{Holdom:1985ag,Holdom:1986eq}. The kinetic mixing between the ordinary photon and this dark photon can endow particles in the hidden sector with a tiny effective electric charge and make them appear millicharged in our detectors.
The existence of millicharged particles could have significant implications. The MCPs are potential candidates for explaining some of the dark matter in the universe~\cite{Goldberg:1986nk,Cheung:2007ut,Feldman:2007wj}, as they would interact very weakly with ordinary matter. They could also subtly affect precision measurements in particle physics and cosmology~\cite{Dobroliubov:1989mr,deMontigny:2023qft,Davidson:2000hf,Dolgov:2013una}.

In this work, we investigate the potential to search for and constrain light MCPs via strong electromagnetic fields in both laboratory laser experiments and astrophysical environments such as magnetars.
The laser pulses of high-intensity enable the study of strong-field physics in terrestrial experiments and have a lot of applications in atomic physics, nuclear physics and particle physics (see recent reviews~\cite{Hartin:2018egj,Fedotov:2022ely} and references therein).
In the 1990s, through the interaction of
an ultra-relativistic electron beam with a terawatt laser pulse, the E144 experiment performed at the Stanford Linear Accelerator Center (SLAC) observed two strong-field processes, i.e., the nonlinear Compton scattering and the nonlinear Breit-Wheeler pair production~\cite{Burke:1997ew,Bamber:1999zt}.
New massive dark particles beyond the SM can also be produced from the laser-assisted nonlinear Compton scattering~\cite{Dillon:2018ypt,King:2018qbq,Dillon:2018ouq,Ma:2024ywm,Ma:2025axq,Li:2025ptt}.
We consider the MCP pair production from laser-assisted nonlinear Compton scattering
\begin{eqnarray}
e^- (+\gamma_{\rm Laser}) \to e^- + \gamma^\ast\to e^-+\chi +\overline{\chi}\;.
\end{eqnarray}
In the presence of an external classical laser field, all charged fermions in the initial and final states can be treated as ``dressed'' states inherently accounting for the continuous interaction with the coherent background of laser photons. For describing the dressed electrons and MCPs, we use the Furry picture and Volkov solution~\cite{Wolkow1935} of Dirac equation in a background electromagnetic field consisting of two polarized plane electromagnetic waves. This trident process is mediated by a virtual photon coupled to MCP pairs. We employ the helicity amplitude method to decompose the total amplitude into two Lorentz invariant amplitudes of production and decay. The cross sections of nonlinear Compton scattering to MCPs will be calculated in terms of both density matrices.
Moreover, the charged and
neutral weak currents mediated by $W^\pm/Z$ bosons in the SM can also result in missing neutrinos in final
states. We take into account the irreducible SM background $e^-\to e^- + \nu\overline{\nu}$.
We finally obtain the sensitivity reach for MCP electric charge fraction $\epsilon$ using laser-assisted nonlinear Compton scattering.

In addition, astrophysical objects such as magnetars have ultra-strong magnetic field in a range of $10^{14}\sim 10^{15}~{\rm G}$ or even beyond. The stellar rotation of magnetars in the
presence of an ultra-strong magnetic field generates a parallel electric field exceeding $10^{12}~{\rm V/m}$ in the polar gap, as demonstrated in the Ruderman-Sutherland (RS) magnetar model~\cite{Ruderman:1975ju}. Although this electric field remains well below the Schwinger limit, its strength significantly exceeds that of laboratory laser fields. If the electric field strength of magnetar is larger than the critical value for MCPs $\mathcal{E}_{\rm cri.}=m_\chi^2/q_\chi$, MCPs with mass much smaller than electron can be spontaneously produced via the Schwinger mechanism from the decay of the electromagnetic vacuum~\cite{Hook:2017vyc,Korwar:2017dio,Kouvaris:2025tom}. After being produced in the polar gap, the Lorentz force from the parallel electric field will accelerate MCPs and result in energy loss with their outflow along open magnetic field lines. This non-perturbative production of MCPs should not alter conventional magnetar nature~\cite{Korwar:2017dio}. In this work, we revisit this energy loss mechanism after
examining the properties of confirmed magnetars based on Ruderman-Sutherland model. We evaluate the energy loss due to the Schwinger pair production of MCPs and show the constraints on the MCP electric charge fraction. The constraints from highly magnetized magnetars and the search potential in laboratory laser experiments are complementary.

This paper is organized as follows. In Sec.~\ref{sec:LaserCompton}, we discuss the production of MCP pairs from laser-assisted nonlinear Compton scattering. We present the strong-field QED framework and the decay widths of the relevant Compton scattering to MCP pairs. In Sec.~\ref{sec:Magnetar}, we consider the Schwinger production of MCP pairs in a strong field
of highly magnetized magnetar. The energy loss induced by MCP Schwinger production from magnetars will be calculated. The sensitivity reach and the constraint on MCP electric charge fraction are shown in Sec.~\ref{sec:results}. Our conclusions are drawn in Sec.~\ref{sec:Con}.

\section{MCP pair production from laser-assisted nonlinear Compton scattering}
\label{sec:LaserCompton}

The wave function of a relativistic fermion with mass $m$ in an electromagnetic potential is governed by the following Dirac equation
\bee
(i \slashed{\partial} - Qe \slashed{A} -m) \psi(x)=0 \;,
\ene
where $e$ is the unit of electric charge and $Q$ is the charge operator (e.g. $Q\psi=-\psi$ for an electron or $Q\psi=\epsilon\psi$ for a MCP).
We assume that the electromagnetic potential $A^\mu(x)$ of the incoming laser field
moves along the direction given by the wave vector $\boldsymbol{k}$ with the on-shell condition $k^2=0$.
To be specific, the laser wave is taken to be circularly polarized and monochromatic~\cite{King:2018qbq}. In the Lorentz gauge $k \cdot A=0$, the vector potential $A^\mu$ can be given as
\begin{equation}
A^\mu(x)=a\left(\varepsilon_1^\mu \cos\phi_x +\varepsilon_2^\mu \sin\phi_x\right) \,,
\end{equation}
where the phase $\phi_x$ is defined as
$\phi_x \equiv k \cdot x = \omega t-\boldsymbol{k} \cdot \boldsymbol{x}$
with $\omega$ being the frequency of the incoming laser,
$\varepsilon_1$ and $\varepsilon_2$ are the two mutually orthogonal polarization vectors. The amplitude $a$ is defined as $a=\mathcal{E}/\omega$ with $\mathcal{E}$ being the strength of the electromagnetic field. It is related to the strength of the laser beam
by the power density
\begin{equation}
I= \frac{1}{4 \pi} a^2 \omega^2 \,.
\end{equation}
One also defines a dimensionless intensity parameter
\begin{eqnarray}
\eta\equiv {ea\over m_e}={e \mathcal{E}\over \omega m_e}\;.
\end{eqnarray}
For the green light with $\omega=2.35$ eV, $\eta=1$ corresponds to $\mathcal{E}\approx 6.1\times 10^{10}~{\rm V/cm}$ and $I\approx 7.8\times 10^{17}~{\rm W}/{\rm cm}^2$.
The above model of highly focused light pulses may be simplified, but the essential properties of the laser field are properly taken into account. In this work, we will use this simplified model to investigate the laser-assisted Compton scattering and the production of MCP pairs. 
For a realistic experimental setup, one can introduce a Gaussian function $f(x)$ to describe the spatial dependence of
the vector potential $A^\mu$ for the pulse shape~\cite{Dillon:2018ypt}. The Fourier transform of the pulse shape $\tilde{f}$ can enter the following calculation of $S$ matrix element.

Since the laser field is taken as a classical external potential,
the higher-order nonlinear effects of the electron decay in the laser field need to be included.
This can be addressed by employing the Volkov state~\cite{Wolkow1935}
which is the exact solution of the above Dirac equation in the presence of a circularly polarized laser field.
For the incoming Dirac particle with $Q$ charge and anti-particle with $-Q$ charge, the wave functions of their Volkov states normalized to the volume $V$ are respectively given as follows~\cite{Wolkow1935}
\begin{eqnarray}
\psi_{p, s}(x)
&=&
\left[1+\frac{Qe \slashed{k} \slashed{A}}{2k \cdot p}\right]
\frac{u\left(p, s\right)}{\sqrt{2 q^0 V}} e^{i F(q, s,x)}\;,\\
\overline{\psi}^{(+)}_{p, s}(x)
&=&
\frac{\overline{v}\left(p, s\right)}{\sqrt{2 q^0 V}}\left[1-\frac{Qe \slashed{k} \slashed{A}}{2k \cdot p}\right]
 e^{i F(q, s,x)}\;,
\end{eqnarray}
where the label ``$(+)$'' refers to the wave function of anti-particle, $u(p, s)$ and $v(p, s)$ are the usual Dirac spinors for the free particle and anti-particle, respectively, and $p$ denotes the initial momentum of Dirac particles before entering the electromagnetic background.
In the external plane-wave field, they gain an ``effective momentum''
\bee
q^\mu
=
p^\mu + \frac{Q^2 e^2 a^2}{2 k \cdot p} k^\mu
\ene
which satisfies the following dispersion relation
\bee
q^2 = m^2 + Q^2 e^2 a^2 = m^{\ast 2}\;.
\ene
The phase function $F(q, s,x)$ is given by
\bee
F(q, s,x)
=
-q \cdot x - \frac{Qea\left(\varepsilon_1 \cdot p\right)}{k \cdot p} \sin \phi_x
+
\frac{Qea\left(\varepsilon_2 \cdot p\right)}{k \cdot p} \cos \phi_x \;.
\ene
The wave functions of the final Volkov states in the laser field can be obtained with the substitution of momentum.
The Volkov wave function as the solution of the Dirac equation describes the ``dressed'' fermion which continuously interacts with the traveling cloud
of laser photons. It includes the nonlinear effects of the laser field in the series of $ea/m$.

\subsection{Nonlinear Compton scattering to MCPs}

Next we consider MCPs ($\chi$ and $\overline{\chi}$) as dark particles produced in the following Compton scattering process
\begin{eqnarray}
e^-(p_1) (+\gamma_{\rm Laser}(k)) \to e^- (p_2) + \chi (p_3) + \overline{\chi}(p_4)\;,
\end{eqnarray}
where the optical photon in laser field has energy $\omega$. This process is mediated by a virtual photon.
The lowest-order scattering $S$ matrix element for the laser-induced MCP production reads
\begin{eqnarray}
S_{f i}
&=&{1\over 2!}
\Big[(ie)^2\epsilon \int d^4x d^4y \overline{\psi}_{p_2,s_2}(x)\gamma_\mu \psi_{p_1,s_1}(x) G^{\mu\nu}(x-y) \overline{\psi}_{p_3,s_3}(y)\gamma_\nu \psi^{(+)}_{p_4,s_4}(y)+x\leftrightarrow y\Big]\;,
\end{eqnarray}
where $G^{\mu\nu}(x-y)=\int {d^4 k'\over (2\pi)^4}{-ig^{\mu\nu}\over k^{\prime 2}} e^{ik'\cdot (x-y)}$ denotes the virtual photon propagator with momentum $k'$.
After employing the Volkov wave functions for both electrons and MCPs, we have
\begin{eqnarray}
S_{fi}={i\over 2!}\Big[\int d^4x d^4y {d^4k'\over (2\pi)^4} {1\over \sqrt{2^4 q_1^0 q_2^0 q_3^0 q_4^0 V^4}} e^{i(q_2-q_1+k')\cdot x} e^{i(q_3+q_4-k')\cdot y}e^{-i\Phi_x} e^{-i\Phi_y}\mathcal{M} + x\leftrightarrow y \Big]\;,
\end{eqnarray}
where $\mathcal{M}$ denotes the matrix element which will be discussed below, and the corresponding phases are given as
\begin{eqnarray}
\Phi_x
&=&
e a\left(\frac{\varepsilon_1 \cdot p_2}{k \cdot p_2}-\frac{\varepsilon_1 \cdot p_1}{k \cdot p_1}\right) \sin \phi_x
-
e a\left(\frac{\varepsilon_2 \cdot p_2}{k \cdot p_2}-\frac{\varepsilon_2 \cdot p_1}{k \cdot p_1}\right) \cos \phi_x\;,\\
&=&ea(\varepsilon_1\cdot \ell \sin\phi_x - \varepsilon_2\cdot \ell \cos\phi_x)\;,
\end{eqnarray}
with $\ell^\mu=p_2^\mu/(k\cdot p_2) - p_1^\mu/(k\cdot p_1)$ and
\begin{eqnarray}
\Phi_y&=&-\epsilon e a\left(\frac{\varepsilon_1 \cdot p_3}{k \cdot p_3}-\frac{\varepsilon_1 \cdot p_4}{k \cdot p_4}\right) \sin \phi_y
+\epsilon
e a\left(\frac{\varepsilon_2 \cdot p_3}{k \cdot p_3}-\frac{\varepsilon_2 \cdot p_4}{k \cdot p_4}\right) \cos \phi_y\;,\\
&=&-\epsilon ea (\varepsilon_1\cdot \ell' \sin\phi_y -
\varepsilon_2\cdot \ell' \cos\phi_y)\;,
\end{eqnarray}
with $\ell^{\prime \mu}=p_3^\mu/(k\cdot p_3) - p_4^\mu/(k\cdot p_4)$. The two phases can then be rewritten as
\begin{eqnarray}
\Phi_x=z\sin(\phi_x-\phi_{0})
\end{eqnarray}
with
\bee
z = ea \sqrt{ (\varepsilon_1 \cdot \ell)^2 + (\varepsilon_2\cdot \ell)^2 } \,,
\quad
\cos\phi_0 =  \frac{e a \varepsilon_1 \cdot \ell }{ z} \,,
\quad
\sin\phi_0 =  \frac{e a \varepsilon_2 \cdot \ell}{ z} \,,
\ene
and
\begin{eqnarray}
\Phi_y=-\epsilon z'\sin(\phi_y-\phi'_{0})
\end{eqnarray}
with
\bee
z' = ea \sqrt{ (\varepsilon_1 \cdot \ell')^2 + (\varepsilon_2\cdot \ell')^2 } \,,
\quad
\cos\phi'_0 =  \frac{e a \varepsilon_1 \cdot \ell' }{ z'} \,,
\quad
\sin\phi'_0 =  \frac{e a \varepsilon_2 \cdot \ell'}{ z'} \,.
\ene
Since $e^{-i\Phi_x}$ is a periodic function, it can be performed by Fourier series expansion as
\bee
e^{-i\Phi_x}
=
e^{- i z \sin \left(\phi_x-\phi_0\right)}
=
\sum_{n=-\infty}^{\infty} c_n e^{-\mathrm{i} n\left(\phi_x-\phi_0\right)}\;,
\ene
where the expansion coefficients are given as
\bee
c_n
=
\frac{1}{2 \pi} \int_{-\pi}^\pi \mathrm{d} \varphi \, e^{-\mathrm{i} z \sin\varphi} e^{\mathrm{i} n \varphi}
=J_n(z)\;.
\ene
This is the integral representation of the Bessel function $J_n(z)$. Thus, we have
\bee
e^{-i\Phi_x} = \sum_{n=-\infty}^{\infty} B_n(z) e^{-\mathrm{i} n \phi_x}\;,
\ene
where $B_n(z)=J_n(z) e^{i n \phi_0}$. This is a well-known relation known as Jacobi-Anger expansion.
Similarly, we get the following ansatz
\begin{eqnarray}
e^{-i\Phi_y}=\sum_{r=-\infty}^{\infty} B_{r}(-\epsilon z') e^{-ir\phi_y}\;,
\end{eqnarray}
where $B_{r}(-\epsilon z')=J_{r}(-\epsilon z')e^{ir\phi'_0}$.
The $S$ matrix element then becomes
\begin{eqnarray}
S_{fi}&=&{i\over 2!}\Big[\int d^4x d^4y {d^4k'\over (2\pi)^4} {1\over \sqrt{2^4 q_1^0 q_2^0 q_3^0 q_4^0 V^4}} \nonumber\\
&&\sum_{n,r} e^{i(q_2-q_1+k'-nk)\cdot x} e^{i(q_3+q_4-k'-rk)\cdot y}B_n(z) B_r(-\epsilon z')\mathcal{M}+x\leftrightarrow y\Big]\;,\\
&=&i\sum_{n,r} {1\over \sqrt{2^4 q_1^0 q_2^0 q_3^0 q_4^0 V^4}} (2\pi)^4 \delta^4 (q_2-q_1+q_3+q_4-nk-rk)B_n(z) B_r(-\epsilon z')\mathcal{M}\;,
\end{eqnarray}
where $k'=q_1+nk-q_2=q_3+q_4-rk$.

The total amplitude can be written as a contraction of two vector currents
\bee
\calm = \frac{\epsilon e^2}{ m_{k'}^2 } g^{\mu\nu} \calm^P_{\mu} \calm^D_{\nu} \,,
\ene
with $\calm^P_{\mu}$ and $\calm^D_{\nu}$ being the currents
for the production and decay of a fictitious spin-1 particle with momentum $k'$ and invariant mass $m_{k'}$,
respectively
\bea
\calm^P_{\mu}
& = &
\Big[ \overline{u}(p_2)
\left(1-\frac{e \slashed{A} \slashed{k} }{2 k \cdot p_2}\right)
\gamma_\mu
\left(1- \frac{e \slashed{k} \slashed{A} }{2 k \cdot p_1}\right) u(p_1)
\Big] \,,
\\[3mm]
\calm^D_{\nu}
&=&
\Big[
\overline{u}(p_3)\left(1+{\epsilon e\slashed{A} \slashed{k}\over 2k\cdot p_3}\right) \gamma_\nu \left(1-{\epsilon e \slashed{k} \slashed{A}\over 2k\cdot p_4}\right) v(p_4)
\Big] \;.
\ena
Note that the decay amplitude can be simplified as
\begin{eqnarray}
\calm^D_{\nu}
\approx
\overline{u}(p_3) \gamma_\nu v(p_4)\,,
\end{eqnarray}
in the limit of $\epsilon\ll 1$. This restores to the result using the wave function of free Dirac particle.
The production and decay parts here are not Lorentz invariant individually.
However, by employing the helicity amplitude method, the amplitude can be decomposed into two Lorentz invariant amplitudes along the momentum $k'$.
This can be addressed by inserting the following relation for the metric tensor
\bee
g^{\mu\nu} = \sum_{\lambda=s,0,\pm 1} \eta_\lambda \varepsilon_\lambda^{\mu*} \varepsilon_\lambda^\nu \,,
\ene
where $\eta_s = 1$ and $\eta_{0,\pm1} = -1$,
and $\varepsilon_\lambda^\mu$ are polarization vectors with helicity
$\lambda = s, 0, \pm1$ projected along the momentum $k'$.
One can easily find that the total amplitude can be rewritten as
\bee
\calm = \frac{\epsilon e^2}{ m_{k'}^2 }\sum_{\lambda=s,0,\pm 1} \eta_\lambda \mathcal{M}_{\lambda}^P \mathcal{M}_{\lambda}^D \,,
\ene
where $\calm^P_{\lambda} = \varepsilon_{\lambda}^{\ast} \cdot \calm^P$ and
$\calm^D_{\lambda} = \varepsilon_{\lambda} \cdot \calm^D$.
It is clear that both $\calm^P_{\lambda}$ and $\calm^D_{\lambda}$
are invariant under Lorentz boost along the direction of the momentum $k'$.
By virtue of this,
the decay amplitude can be calculated in the rest frame of the momentum $k'$.
We define
\begin{eqnarray}
\calm^P_{\lambda,n}&=&B_n(z)\calm^P_{\lambda}\;,\\
\calm^D_{\lambda,r}&=&B_r(-\epsilon z') \calm^D_{\lambda}\;,\\
\calm_{n,r}&=&B_n(z) B_r(-\epsilon z') \calm=\frac{\epsilon e^2}{ m_{k'}^2 }\sum_{\lambda=s,0,\pm 1} \eta_\lambda \mathcal{M}_{\lambda,n}^P \mathcal{M}_{\lambda,r}^D\;.
\end{eqnarray}
The $S$ matrix then becomes
\begin{eqnarray}
S_{fi}&=&i\sum_{n,r} {1\over \sqrt{2^4 q_1^0 q_2^0 q_3^0 q_4^0 V^4}} (2\pi)^4 \delta^4 (q_2-q_1+q_3+q_4-nk-rk)\nonumber\\
&&\times \frac{\epsilon e^2}{ m_{k'}^2 }\sum_{\lambda=s,0,\pm 1} \eta_\lambda \mathcal{M}_{\lambda,n}^P \mathcal{M}_{\lambda,r}^D\;.
\end{eqnarray}
When this $S$ matrix is squared, one has to deal with
\begin{eqnarray}
\delta^4(q_2-q_1+q_3+q_4-nk-rk)\delta^4(q_2-q_1+q_3+q_4-n'k-r'k)\;.
\end{eqnarray}
In the following calculation, we take $n>0$ for the absorption of at least one laser photon by electron beam and $r=0$ to include the leading order contribution in the production of MCP pairs. As a result, the interference terms in the summation of $n$ vanish due to the product of two delta functions.
After squaring the $S$ matrix and taking into account the corresponding 3-body phase space, the laser-assisted (L) decay width of the incoming dressed electron is given as
\bee
\Gamma_{\chi\overline{\chi}}^{\rm L} = \sum_{n=1}^{\infty} \frac{1}{2Q_e} \int d\Pi_{P,n} \int d\Pi_{D} \int \frac{dm_{k'}^2}{2\pi} \left( \frac{\epsilon e^2}{m_{k'}^2} \right)^2 \sum_{\lambda, \lambda' = s,0,\pm 1} \eta_\lambda \eta_{\lambda'} \mathcal{P}_{\lambda\lambda',n} {\mathcal{D}_{\lambda\lambda',r=0}} \,,
\ene
where $Q_e$ is the energy of the incoming dressed electron in the laboratory frame, $d\Pi_{P,n}$ and $d\Pi_{D}$ denote the phase space of the production $d\Pi_2(q_1+nk-q_2-k')$ and the decay $d\Pi_2(k'-q_3-q_4)$, respectively.
The spin of the incoming electron has been averaged, and the spins of the outgoing MCPs have been summed over implicitly. The density matrices of production and decay are
\begin{eqnarray}
{\mathcal{P}_{\lambda\lambda',n}}
&=&{1\over 2} \mathcal{M}^P_{\lambda,n}(\mathcal{M}_{\lambda',n}^{P})^\dagger\nonumber \\
&=& \delta_{\lambda\lambda'}
    \begin{cases}
        0 \,, & \text{for } \lambda = s \\[2ex]
        -\frac{4}{3} J_n^2 \left( m_e^2 + \frac{1}{2} m_{k'}^2 \right) - \frac{4}{3}e^2 a^2 \mathcal{J}_n \left( u + \frac{1}{4u} \right)  \,, & \text{for } \lambda = 0, \pm 1
    \end{cases}
\end{eqnarray}
where
\bee
\mathcal{J}_n = J_n^2 - \frac{1}{2} \left( J_{n-1}^2 + J_{n+1}^2 \right) \,,~~~u={k\cdot q_1\over 2k\cdot q_2}\;,
\ene
and
\begin{eqnarray}
{\mathcal{D}_{\lambda\lambda',r}}
&=&\mathcal{M}^D_{\lambda,r}(\mathcal{M}_{\lambda',r}^{D})^\dagger\nonumber \\
&=& \delta_{\lambda\lambda'}
    \begin{cases}
        0 \,, & \text{for } \lambda = s \\[2ex]
        \frac{8}{3} J_r^2 \left( m_\chi^2 + \frac{1}{2} m_{k'}^2 \right) - \frac{8}{3}\epsilon^2e^2 a^2 \mathcal{J}_r \left( u' + \frac{1}{4u'} \right)  \,, & \text{for } \lambda = 0, \pm 1
    \end{cases}
\end{eqnarray}
where
\bee
\mathcal{J}_r = J_r^2 - \frac{1}{2} \left( J_{r-1}^2 + J_{r+1}^2 \right)\,,~~~u'={k\cdot q_4\over 2k\cdot q_3}\;,
\ene
with $r=0$.
The 2-body phase space can be given in terms of the variable $u$ or $u'$. For instance, the phase space of production is
\bee
d\Pi_{P,n} = \frac{1}{32\pi^2 u^2} du d\phi^\ast_n \,,
\ene
with $\phi^\ast_n$ being the azimuthal angle of the momentum $k'$. The 2-body decay can be calculated in the rest frame of $k'$.

\begin{figure}[htbp]
\centering
\includegraphics[width=0.475\textwidth]{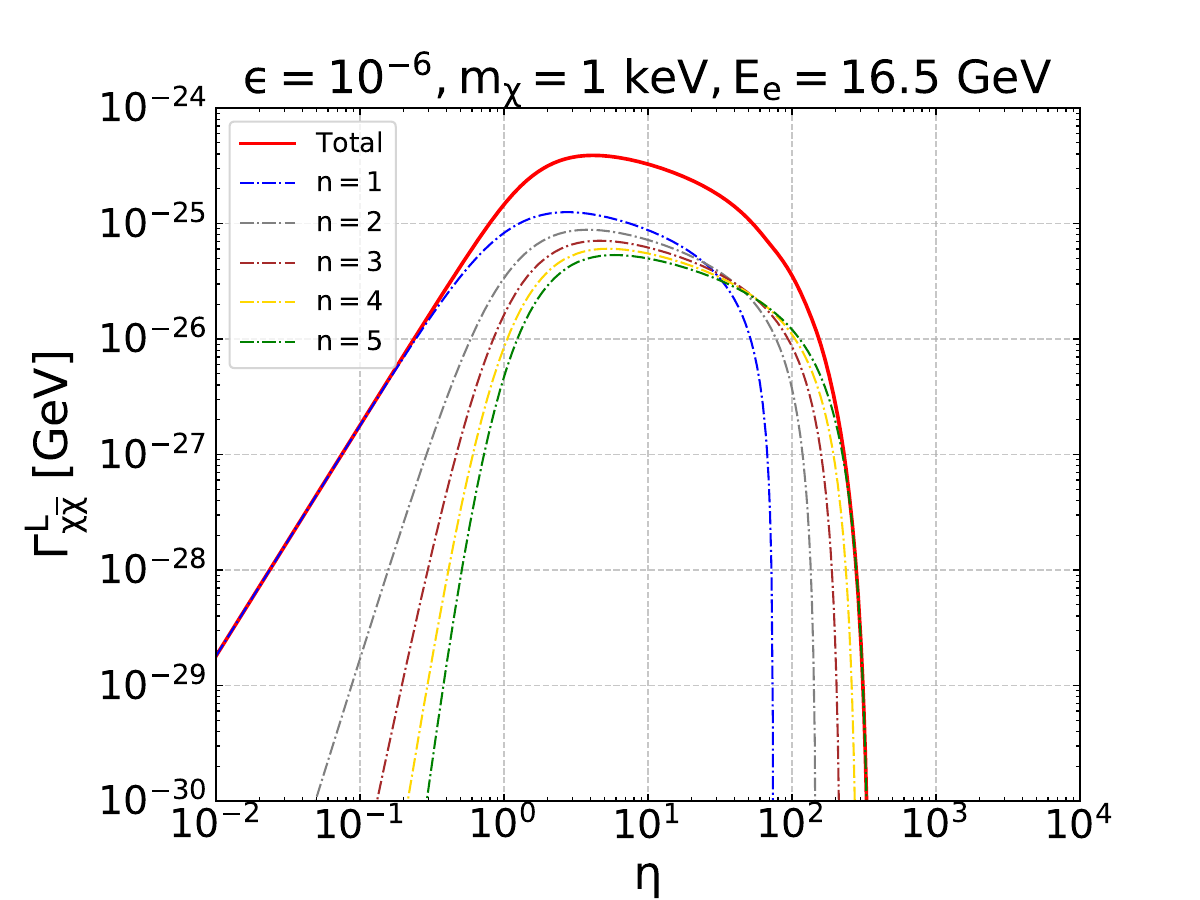}
\includegraphics[width=0.475\textwidth]{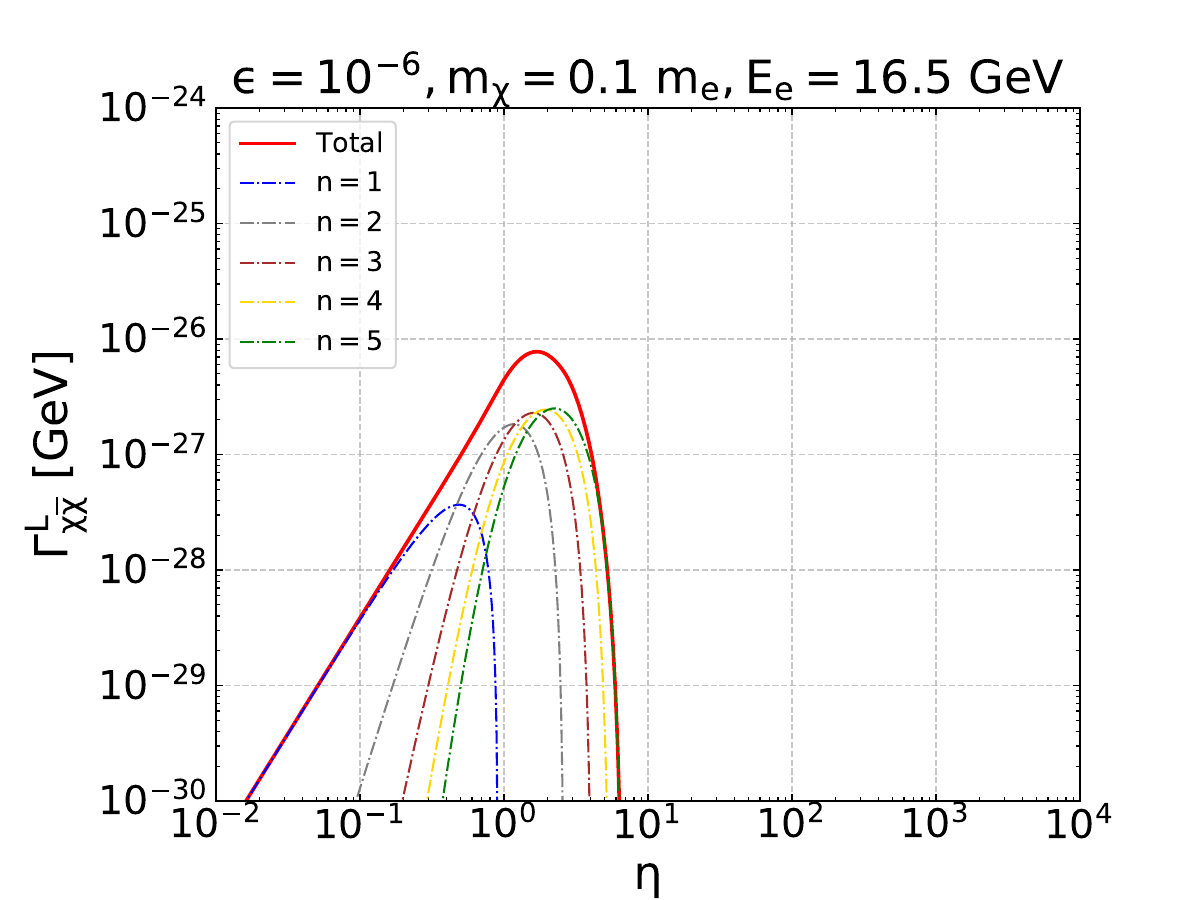}
\includegraphics[width=0.475\textwidth]{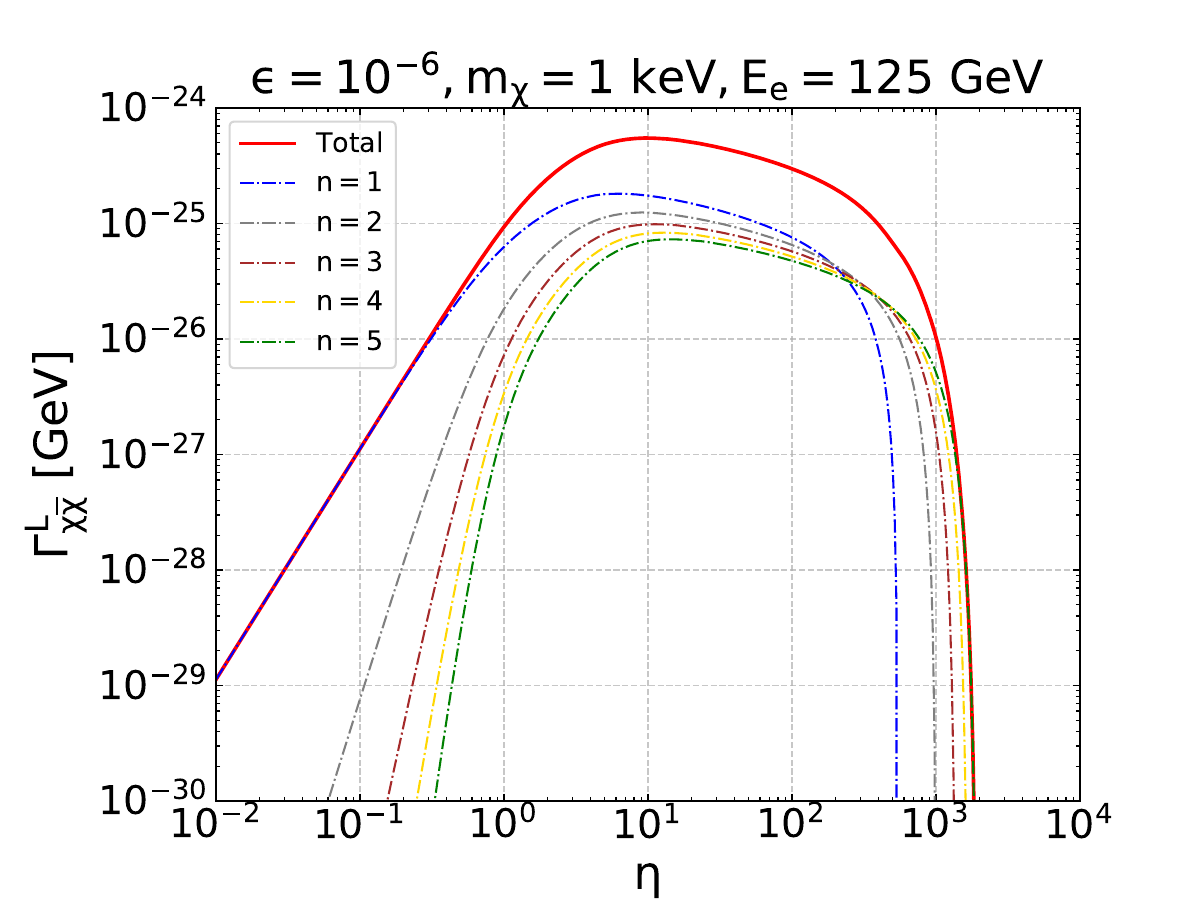}
\includegraphics[width=0.475\textwidth]{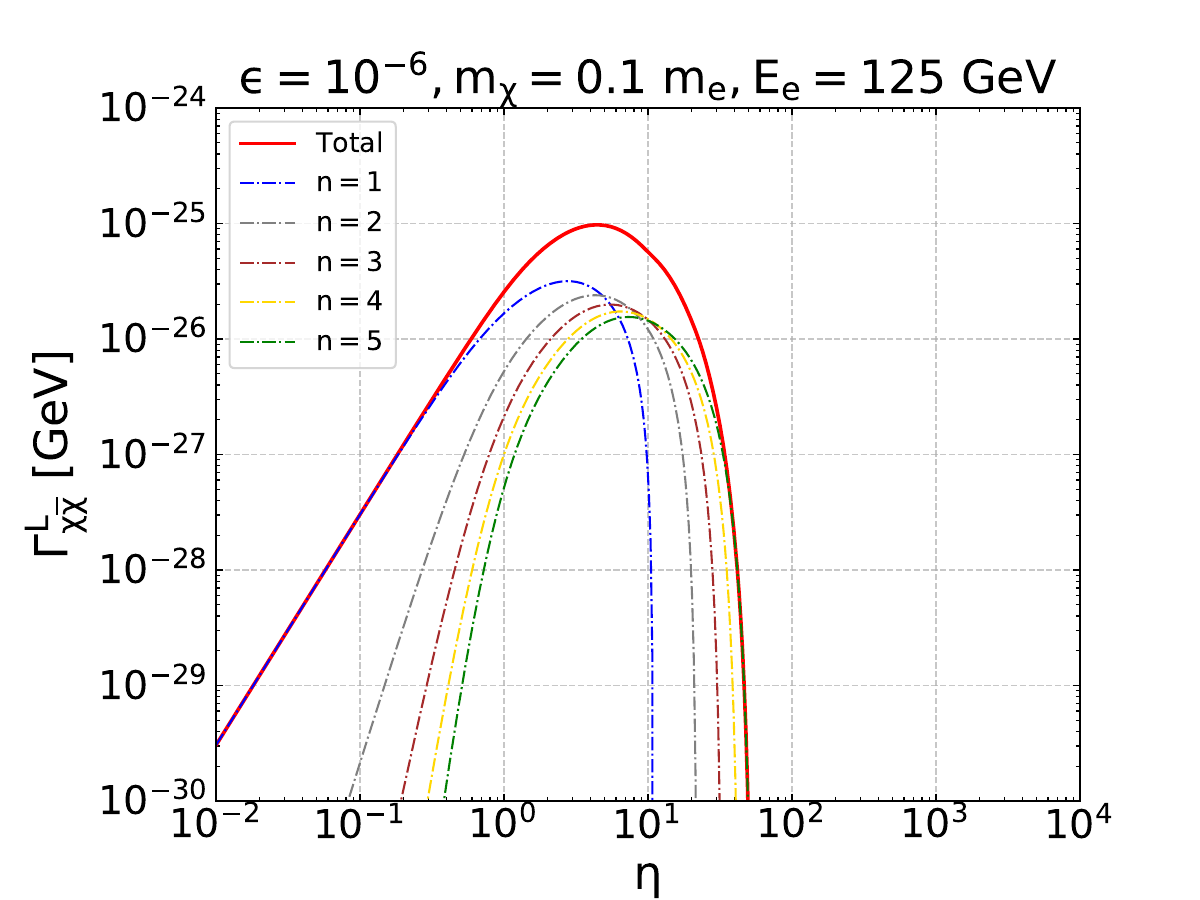}
\caption{The laser-assisted decay widths of the incoming dressed electron as a function of the laser intensity parameter $\eta$
with $\epsilon=10^{-6}$ and $m_\chi=1~{\rm keV}$ (left panels) or $m_\chi=0.1 m_e$ (right panels). We take the energy of electron beam $E_e=16.5$ GeV (upper panels) or $E_e=125$ GeV (lower panels), laser light of energy $\omega=2.35$ eV and the initial scattering angle $\theta=17^\circ$.}
\label{fig:GammaL}
\end{figure}

Fig.~\ref{fig:GammaL} shows the laser-assisted decay widths of the incoming dressed electron $\Gamma_{\chi\overline{\chi}}^{\rm L}$, as a function of the laser intensity parameter $\eta$
with $\epsilon=10^{-6}$ and $m_\chi=1~{\rm keV}$ (left panels) or $m_\chi=0.1 m_e$ (right panels). The energy of electron beam is taken as $E_e=16.5$ GeV (upper panels) or $E_e=125$ GeV (lower panels).
The total widths are shown by solid curves,
and the contributions from the $n$-th branch of laser photon are shown by blue dash-dotted $(n=1)$,
gray dash-dotted $(n=2)$, brown dash-dotted $(n=3)$, yellow dash-dotted $(n=4)$ and green dash-dotted $(n=5)$ curves. One can clearly see that for small values of $\eta$,
the dominant contribution is given by the $n=1$ branch.
When $\eta$ increases, the higher-order absorptions grow quickly and offer significant contributions.
For lighter MCP ($m_\chi = 1$ keV), higher-order contributions can enhance the total decay width by a few times for $\eta\gtrsim 1$.
One can also see that, for heavier MCP ($m_\chi = 0.1m_e$),
the enhancement of higher-order contributions becomes more important.
For every branch, there is a maximum for the intensity parameter $\eta$
\bee
\eta^{\rm max}_n
\simeq
\Big[\Big({2n\omega(E_e+\cos\theta p_e)- 4m_\chi^2\over 4m_e m_\chi}\Big)^2-1\Big]^{1/2}\;,~~~p_e=\sqrt{E_e^2-m_e^2}\;.
\ene
There is no enough phase space for the decay beyond this value. The width dramatically decreases at larger $\eta$ due to this kinematic suppression.

\subsection{Laser-induced SM backgrounds}

Next, we analyze the relevant SM background processes mediated by $W^\pm$ or $Z$ bosons in details. They give the same single fermion final state with missing neutrinos
\begin{eqnarray}
e^-(p_1) (+ \gamma_{\rm Laser} (k)) \to e^-(p_2) + \nu_\ell(p_4) + \overline{\nu_\ell}(p_3)\;,
\end{eqnarray}
where $p_3+p_4=k'$ and $\ell=e,\mu,\tau$.

The charged and neutral weak currents in the SM result in four-fermion effective operators after integrating out the heavy $W^\pm$ and $Z$ bosons.
The relevant four-fermion effective Lagrangian for neutrino-electron interaction is
\begin{eqnarray}
-\mathcal{L}^{\nu_\ell e}&=&{G_F\over \sqrt{2}} \overline{\nu}_\ell\gamma^\mu (1-\gamma_5)\nu_\ell \overline{e}\gamma_\mu(g_{LV}^{\nu_\ell e}-g_{LA}^{\nu_\ell e} \gamma_5)e\;,
\end{eqnarray}
where
\begin{eqnarray}
g_{LV}^{\nu_e e} &=& {1\over 2} + 2\sin^2\theta_W\;,\\
g_{LA}^{\nu_e e} &=& {1\over 2}\;,\\
g_{LV}^{\nu_\ell e} &=& -{1\over 2} + 2\sin^2\theta_W\;,~~~\nu_\ell=\nu_\mu, \nu_\tau\\
g_{LA}^{\nu_\ell e} &=& -{1\over 2}\;,~~~\nu_\ell=\nu_\mu, \nu_\tau\;.
\end{eqnarray}
The $S$ matrix element for the production of neutrinos from the collision between laser and electron beam becomes
\begin{eqnarray}
S_{fi}^{\nu_\ell e} &=&-i\frac{G_F}{\sqrt{2}}\int d^4x  \Big(\overline{\psi}_{p_2, s_2}(x) \gamma_\mu(g_{LV}^{\nu_\ell e} -g_{LA}^{\nu_\ell e}\gamma_5) \psi_{p_1, s_1}(x)\Big)  \Big(\overline{\psi}_{p_4, s_4}(x) \gamma^\mu(1-\gamma_5)  \psi_{p_3, s_3}(x)\Big) \nonumber\\
&=& -i\frac{G_F}{\sqrt{2}} \int d^4 x~ e^{-i(q_1-q_2-p_3-p_4) \cdot x} \mathcal{M}^{\nu_\ell e}~e^{-i\Phi_x}\;,
\end{eqnarray}
where the wave functions of dressed electron and free neutrinos are
\begin{eqnarray}
\psi_{p_1, s_1}(x) & = & \left[ 1 - \frac{e\slashed{k}\slashed{A}}{2k\cdot p_1} \right] u(p_1,s_1) ~e^{iF(q_1,s_1)}\;, \\[0.5em]
\overline{\psi}_{p_2, s_2}(x) & = & \overline{u}(p_2, s_2)  ~e^{-i F(q_2, s_2)} \left[ 1 - \frac{e\slashed{A}\slashed{k}}{2k\cdot p_2} \right]\;, \\[0.5em]
\psi_{p_3, s_3}(x) & = & v(p_3,s_3) ~e^{ip_3 \cdot x}\;, \\[0.5em]
\overline{\psi}_{p_4, s_4}(x) & = & \overline{u}(p_4, s_4) ~e^{i p_4 \cdot x}\;.
\end{eqnarray}
For this calculation, we also use the method of helicity amplitude in terms of the density matrices of production and decay parts.
The total amplitude can be written as a product of two currents
\begin{eqnarray}
\mathcal{M}^{\nu_\ell e} &=& \Big[ \bar{u}(p_2,s_2)(1 - \frac{e\slashed{A}\slashed{k}}{2k\cdot p_2})\gamma_\mu(g_{LV}^{\nu_\ell e} -g_{LA}^{\nu_\ell e}\gamma_5)( 1 - \frac{e\slashed{k}\slashed{A}}{2k\cdot p_1})u(p_1,s_1)\Big]\Big[ \bar{u}(p_4,s_4)\gamma^\mu(1-\gamma_5)u(p_3,s_3)\Big] \nonumber \\[1em]
&=& g^{\mu\nu}\mathcal{M}_{P,\mu}^{\nu_\ell e} \mathcal{M}_{D,\nu}^{\nu_\ell e}\;,
\end{eqnarray}
where the amplitudes $\mathcal{M}_{P}^{\nu_\ell e}$ and $\mathcal{M}_{D}^{\nu_\ell e}$ describe the production and decay of a fictitious spin-one particle
with momentum $k'$, respectively, and both of them are Lorentz invariant.
After employing the helicity amplitude method, this amplitude can be decomposed into two Lorentz invariant amplitudes along the momentum $k'$
\begin{eqnarray}
\mathcal{M}^{\nu_\ell e} &=&\sum_{\lambda=s,0,\pm 1} \eta_\lambda \mathcal{M}_{P,\lambda}^{\nu_\ell e} \mathcal{M}_{D,\lambda}^{\nu_\ell e}\;,
\end{eqnarray}
where $\lambda=s,0,\pm 1$ is the helicity projected
along the momentum $k'$, and $\mathcal{M}_{P,\lambda}^{\nu_\ell e}=\varepsilon_\lambda^\ast \cdot \mathcal{M}_{P}^{\nu_\ell e}$ and $\mathcal{M}_{D,\lambda}^{\nu_\ell e}=\varepsilon_\lambda \cdot \mathcal{M}_{D}^{\nu_\ell e}$ with $\varepsilon_\lambda^{(\ast)\mu}$ being the polarization vectors.

By considering the absorption of $n$ laser photons, the $S$ matrix can be rewritten as
\begin{eqnarray}
S_{fi}^{\nu_\ell e}&=&-i\frac{G_F}{\sqrt{2}} \sum_{n=-\infty}^{\infty} \int d^4 x~ e^{-i(q_1+nk-q_2-p_3-p_4) \cdot x} \mathcal{M}^{\nu_\ell e}_n(z)\;,
\end{eqnarray}
where the $n$-th amplitude is $\mathcal{M}^{\nu_\ell e}_n(z)=\mathcal{M}^{\nu_\ell e}_{P,n}(z)\cdot \mathcal{M}^{\nu_\ell e}_{D}$ with $\mathcal{M}^{\nu_\ell e}_{P,n}(z)=B_n(z)\mathcal{M}^{\nu_\ell e}_{P}$. Then, the laser-assisted decay width becomes
\begin{eqnarray}
\Gamma_{\nu\overline{\nu}}^{\rm L}&=&\frac{G_F^2}{2} \sum_{\ell=e,\mu,\tau}\sum_{n=1}^{\infty} \frac{1}{2Q_e} \int d\Pi_{P,n} \int \frac{dm_{k'}^2}{2\pi} \sum_{\lambda,\lambda'=s,0,\pm 1} \eta_\lambda \eta_{\lambda'} \mathcal{P}_{\lambda\lambda',n}^{\nu_\ell e} \overline{\mathcal{D}_{\lambda\lambda'}^{\nu_\ell e}}\;,
\end{eqnarray}
where
\begin{eqnarray}
d\Pi_{P,n} = \frac{1}{32\pi^{2}u^{2}}  du  d\phi_{n}^{*}
\end{eqnarray}
with $\phi_{n}^{*}$ being the azimuthal angle of the momentum $k'$. After integrating out the two-body phase space
of the outgoing SM neutrinos, decay density matrices can be easily calculated. The non-zero density matrices of decay and production parts are
\begin{eqnarray}
\overline{\mathcal{D}_{ss}^{\nu_\ell e}}  &=& \displaystyle\frac{m_{k^\prime}^2}{4\pi}  \beta \left( 1 -  \beta^{2} \right) \;,~~\beta=\sqrt{1-4m_\nu^2/m_{k'}^2}\approx 1\nonumber \\[0.5em]
\overline{\mathcal{D}_{\lambda\lambda}^{\nu_\ell e}} & =& \displaystyle\frac{m_{k^\prime}^2}{4\pi}  \beta \left( 1 + \frac{1}{3} \beta^{2} \right), ~~~~\text{for } \lambda = 0, \pm 1
\end{eqnarray}
and
\begin{eqnarray}
\mathcal{P}_{ss,n}^{\nu_\ell e} & = & 4J_n^2~(g_{LA}^{\nu_\ell e})^2m_e^2 +8e^2a^2\mathcal{J}_n~ (g_{LA}^{\nu_\ell e})^2\frac{m_e^2}{m_{k^\prime}^2}u(1-\frac{1}{2u})^2\;, \nonumber \\[0.5em]
\sum_{\lambda=0,\pm1} \mathcal{P}_{\lambda\lambda,n}^{\nu_\ell e} & = & -2 J_n^2~\Big[(g_{LV}^{\nu_\ell e})^2(m_{k^\prime}^2+2m_e^2)+(g_{LA}^{\nu_\ell e})^2(m_{k^\prime}^2-4m_e^2) \Big] \nonumber \\
&& - 4e^2a^2\mathcal{J}_n ~\Big[ (g_{LV}^{\nu_\ell e})^2(u+\frac{1}{u})+(g_{LA}^{\nu_\ell e})^2\Big( (u+\frac{1}{4u})+2\frac{m_e^2}{m_{k^\prime}^2}u(1-\frac{1}{2u})^2   \Big) \Big]\;, \nonumber
\end{eqnarray}
where $\mathcal{J}_n\equiv J_n^2-{1\over 2}\Big(J_{n-1}^2+J_{n+1}^2\Big)$. In Table~\ref{tab:xsecbkg}, we show the numerical decay width of background $e^-+{\rm Laser}\to e^-+\nu \overline{\nu}$ $\Gamma_{\nu\overline{\nu}}^{\rm L}$ for some $E_e$ and $\eta$ benchmarks. They are negligibly small compared with our signal processes.
Nevertheless, we take into account this background in the following analysis of prospective sensitivity.

In principle, there can be other reducible backgrounds. For instance, the photons in the nonlinear Compton scattering $e^-\to e^-+n\gamma$ may not be recorded by
the detector. When the energy threshold of electron-positron pair is achieved, the trident electron-positron pair production $e^-\to e^- e^+e^-$ would also contribute to backgrounds if the electron-positron pair is misidentified in the detector. However, all these processes are reducible and cannot dominate over the irreducible background. It is sufficient to only consider the irreducible background for a primary estimation of the experimental sensitivity.

\begin{table}[htbp]
\centering
\begin{tabular}{l|c|c|c}
\hline\hline
 $\Gamma_{\nu\overline{\nu}}^{\rm L}$ (GeV) & $\eta=0.5$ & $\eta=10$ & $\eta=100$ \\
\hline
$E_e=16.5$ GeV & $8.64 \times 10^{-38}$ & $2.56 \times 10^{-38}$ & $2.57 \times 10^{-40}$\\
\hline
$E_e=125$ GeV & $3.83 \times 10^{-36}$ & $2.85 \times 10^{-36}$ & $1.49 \times 10^{-38}$\\
\hline\hline
\end{tabular}
\caption{The decay widths of background $e^-+{\rm Laser}\to e^-+\nu \overline{\nu}$. We take $\eta=0.5$, 10 or 100 and electron beam energy $E_e=16.5~{\rm GeV}$ or $E_e=125~{\rm GeV}$.}
\label{tab:xsecbkg}
\end{table}

\section{Energy loss induced by MCP Schwinger production from magnetars}
\label{sec:Magnetar}

In the previous section, we explore the MCP production through laser-assisted nonlinear Compton scattering in an electric field weaker than the critical field $\mathcal{E}_{\rm cri.}=m_\chi^2/q_\chi$ with $q_\chi=\epsilon e$ being the MCP electric charge. If the electric field exceeds this critical value,
\begin{eqnarray}
\mathcal{E}\gtrsim \mathcal{E}_{\rm cri.}={m_\chi^2\over q_\chi}\;,
\label{eq:cri}
\end{eqnarray}
MCPs can be spontaneously produced via the Schwinger effect from the decay of the electromagnetic vacuum.
In this section, we consider the Schwinger production of MCP pairs in a strong field of highly magnetized neutron stars (NSs) and the consequent constraints on the electric charge fraction.

Magnetars as a class of young NSs own ultra-strong magnetic field in a range of $10^{14}\sim 10^{15}~{\rm G}$ or even greater (see a review in Ref.~\cite{Kaspi:2017fwg}).
In 1969, Goldreich and Julian established the fundamental electrodynamics of NS magnetospheres~\cite{Goldreich:1969sb}. They demonstrated that stellar rotation in the presence of a strong magnetic field induces a quasi-static electric field perpendicular to the magnetic field lines.
The generated strong electric field extracts charged particles, such as electrons and protons, from the NS surface. The continuous outflow of charged particles along open magnetic field lines, known as Goldreich-Julian (GJ) current, neutralizes the incipient parallel electric field and forms a force-free magnetosphere in the co-rotating frame around NS.

However, as discussed in the Ruderman-Sutherland model~\cite{Ruderman:1975ju}, the NS surface cannot supply the positive GJ current required to maintain force-free co-rotation in the polar open field line region. For NSs where the product of the spin axis and surface magnetic field $\mathbf{\Omega}\cdot \mathbf{\mathcal{B}}<0$, the required GJ charge density $\rho_{\rm GJ}$ is positive. This demands an outward flow of positive ions to screen the parallel electric field $\mathcal{E}_\parallel$ and keep plasma tied to the rotating magnetic field. However, Ruderman and Sutherland pointed out that heavy ions (dominantly $^{56}{\rm Fe}$ nuclei in the NS crust) are too tightly bound to the surface by cohesive forces because their binding energy exceeds the maximum rotational potential drop available. Thus, positive ions cannot be extracted even under the induced electric field. This shortage of positive charge leaves the local plasma density well below $\rho_{\rm GJ}$ and allows parallel electric field $\mathcal{E}_\parallel$ to accumulate above the surface. This mechanism forms a near-vacuum acceleration region just above the polar cap, i.e., the so-called ``polar gap'' region. The strong $\mathcal{E}_\parallel$ in the gap accelerates particles and emits curvature radiation photons that decay into electron-positron pairs in the strong magnetic field. Once the pair plasma density grows high enough to short out $\mathcal{E}_\parallel$, the gap discharges temporarily. It reforms as the plasma streams away and the positive charge deficit recurs. The cyclic activity is created to drive pulsar emission.

The rotation of magnetars with $\mathcal{O}(1)~{\rm s}$ spin period can induce strong electric field exceeding $10^{12}~{\rm V/m}$. If the strong parallel electric field in the polar gap exceeds the critical strength in Eq.~(\ref{eq:cri}), MCPs with small mass can be produced in pairs via the Schwinger effect from the electromagnetic vacuum~\cite{Hook:2017vyc,Korwar:2017dio,Kouvaris:2025tom}. In a frame in which $\vec{\mathcal{E}}$ and $\vec{\mathcal{B}}$ are parallel (both $\vec{\mathcal{E}}^2-\vec{\mathcal{B}}^2$ and $\vec{\mathcal{E}}\cdot\vec{\mathcal{B}}$ not-zero~\cite{Kim:2003qp}) with magnitude $\mathcal{E}$ and $\mathcal{B}$, respectively, one can obtain the imaginary part of one-loop effective action per four-volume for spinor QED~\cite{Sauter:1931zz,Heisenberg:1936nmg,Weisskopf:1936hya,Schwinger:1951nm,Nikishov:1969tt,BunkinTugov1970,Popov1972,Daugherty:1976mg}
\begin{equation}
2\mathrm{Im}\mathcal{L}_{\rm spinor}=\frac{q_\chi^2 \mathcal{E}\mathcal{B}}{(2\pi)^2}\sum_{n=1}^{\infty}\frac{1}{n}\mathrm{coth}\Big(\frac{n\pi \mathcal{B}}{\mathcal{E}}\Big)~\mathrm{exp}\Big(-\frac{n\pi m^2_\chi}{q_\chi\mathcal{E}}\Big)\;.
\end{equation}
The MCP pair production rate per volume per time in magnetar ${\rm M}$ is the first term in these series
\begin{eqnarray}
\Gamma_{\chi\overline{\chi}}^{\rm M}=\frac{q_\chi^2 \mathcal{E}\mathcal{B}}{(2\pi)^2}\mathrm{coth}\Big(\frac{\pi \mathcal{B}}{\mathcal{E}}\Big)~\mathrm{exp}\Big(-\frac{\pi m^2_\chi}{q_\chi\mathcal{E}}\Big)\;.
\label{eq:pair}
\end{eqnarray}
It turns out that when $\mathcal{E}$ is less than the critical strength $m_\chi^2/q_\chi$, the MCP pair production rate would be essentially suppressed by ${\rm exp}(-{\pi m_\chi^2\over q_\chi \mathcal{E}})$. After being produced in the polar gap, the Lorentz force will accelerate MCPs to the top of gap and induce energy loss along with their outflow. Next, we discuss the properties of magnetars based on RS model and evaluate the energy loss due to the Schwinger pair production of MCPs.

The volume of polar gap as a cylindrical region of maximum height $h$ in the RS model is estimated to be~\cite{Ruderman:1975ju}
\begin{equation}
h \simeq5\times10^3~\mathrm{cm}~ \Big({\rho\over 10^6~{\rm cm}}\Big)^{2/7}\Big({P\over 1~{\rm s}}\Big)^{3/7}\Big({\mathcal{B}\over 10^{12}~{\rm G}}\Big)^{-4/7}\;,
\end{equation}
where $\rho=9\times10^7~{\rm cm} \sqrt{P/(1~{\rm s})}$~\cite{Timokhin:2015dua,Timokhin:2018vdn,Caputo:2023cpv} is the characteristic radius of curvature of the magnetic field line for electron motion, $P$ is the rotation period in units of second, and $\mathcal{B}$ is the radial component of the surface field strength in units of Gauss.
The potential difference across the gap is found to be
\begin{eqnarray}
\Delta V\simeq 1.6\times 10^{12}~\mathrm{V}~ \Big({\rho\over 10^6~{\rm cm}}\Big)^{4/7}\Big({P\over 1~{\rm s}}\Big)^{-1/7}\Big({\mathcal{B}\over 10^{12}~{\rm G}}\Big)^{-1/7}\;.
\end{eqnarray}
The electric field at the bottom of polar gap is
\begin{equation}
\mathcal{E}_{\rm max}\simeq {2\Delta V\over h}\simeq (6.4\times 10^{8}~\mathrm{V/cm})~\Big({\rho\over 10^6~{\rm cm}}\Big)^{2/7}\Big({P\over 1~{\rm s}}\Big)^{-4/7}\Big({\mathcal{B}\over 10^{12}~{\rm G}}\Big)^{3/7}\;.
\end{equation}
Then, the component of the electric field in the gap becomes $\mathcal{E}(z)\simeq \mathcal{E}_{\rm max}(1-z/h)$, which vanishes at the top $z=h$.
The volume of the pulsar gap is given by
\begin{eqnarray}
V&=&\pi(r_p^2-r_c^2)h\;,\\
r_p&=&R_{\rm M}\sqrt{R_{\rm M}/R_{\rm LC}}=\sqrt{2\pi R_{\rm M}^3/P}\;,~~r_c=\Big({2\over 3}\Big)^{3/4}r_p\approx 0.74 r_p\;,
\end{eqnarray}
where $R_{\rm M}$ is the magnetar radius, $R_{\rm LC}=P/(2\pi)$ is the light cylinder radius, $r_p$ is the distance from the magnetic pole to the starting position of the last open magnetic field line which still manages to cross the light cylinder, and $r_c$ is the distance from the pole to the starting position of a special magnetic field line which intersects the light cylinder perpendicularly relative to the magnetic axis. Therefore $r_p$ defines the region of open field lines that can actually emit and the outer boundary of the polar cap, while $r_c$ serves as the inner radius of an annular polar cap with gap height $h$.
Note that the above volume is estimated under the alignment assumption between the rotation axis and the magnetic dipole axis. In this configuration, electron emission occurs near the pole up to $r_c$. This results in a positive current flowing outward through the annulus $r_c<r<r_p$. Conversely, anti-alignment shifts the gap formation to the region within $r_c$. The angle between these axes lies between these two idealized cases in realistic pulsars. We follow Ref.~\cite{Kouvaris:2025tom} to adopt the aligned model for a conservative calculation of the polar gap volume. We adopt a canonical magnetar radius $R_{\rm M}\sim 10~{\rm km}$. This is consistent with the neutron star radius taken in the Australia Telescope National Facility (ATNF) pulsar catalog~\cite{Manchester:2004bp} and that in McGill magnetar catalog~\cite{Olausen:2013bpa,McGillMagnetarCatalog}. In Table~\ref{tab:mcgill}, we show some persistent characteristics relevant for our calculation (rotation period $P$, surface magnetic field strength $\mathcal{B}$ and X-ray luminosity $L_X$) of the confirmed magnetars with $\mathcal{B}>10^{14}~{\rm G}$ in McGill catalog~\cite{McGillMagnetarCatalog} (to the left of the vertical double-line) and the consequent properties of polar gap (to the right of the vertical double-line). One can see that the height of polar gap is at least three orders of magnitude smaller than the magnetar radius. The order of magnitude of maximal electric field strength in the polar gap can reach as large as $\sim 10^{12}~{\rm V/m}$.

\begin{table}[htbp]
\centering
\scalebox{0.9}{
\begin{tabular}{c|c|c|c||c|c|c}
\hline\hline
Name & $P~({\rm s})$ & $\mathcal{B}~(10^{14}~{\rm G})$&$L_X~(10^{33}~{\rm erg\ s^{-1}})$&$h~({\rm m})$ & $V~(10^4~{\rm m^3})$ & $\mathcal{E}_{\rm max}~(10^{11}~{\rm V/m})$\\
\hline
CXOU J010043.1-721134 & 8.020392(9) & 3.9 &65&19.7& 7.35 & 12.2  \\
SGR 0501+4516 &5.7620695(1)&1.9 &0.81&24.5&12.8&10.4 \\
SGR 0526-66 &8.0544(2)&5.6&189&16.0&5.96&14.3 \\
1E 1048.1-5937 &6.457875(3)& 3.9&49&17.4&8.06&13.4 \\
1E 1547.0-5408 ($\checkmark$) &2.0721255(1)&3.2&1.3&10.2&14.7&20.1 \\
PSR J1622-4950 &4.3261(1)&2.7&0.44&17.0&11.8&13.6 \\
SGR 1627-41 &2.594578(6)& 2.2&3.6&14.3&16.5&15.5    \\
1RXS J170849.0-400910 &11.00502461(17)&4.7&42&21.2&5.77&11.6  \\
CXOU J171405.7-381031 &3.825352(4)&5.0&56&11.2&8.75&18.7  \\
SGR 1806-20 ($\checkmark$) &7.54773(2)&20&163&7.46&2.96&25.3\\
XTE J1810-197 ($\checkmark$) &5.5403537(2)&2.1& 0.043&22.7&12.3 &11.0  \\
1E 1841-045 &11.788978(1)&7.0&184&17.5&4.46&13.3   \\
SGR 1900+14 &5.19987(7)&7.0&90&11.0 &6.33&18.9    \\
\hline\hline
\end{tabular}
}
\caption{Some persistent characteristics (rotation period $P$, surface magnetic field strength $\mathcal{B}$, and X-ray luminosity $L_X$) of the confirmed magnetars with $\mathcal{B}>10^{14}~{\rm G}$ in McGill catalog~\cite{McGillMagnetarCatalog} (to the left of the vertical double-line) and the consequent properties of polar gap (to the right of the vertical double-line). The three magnetars with check mark are chosen for later calculation.
}
\label{tab:mcgill}
\end{table}

The Schwinger pair production of MCPs will consume the electromagnetic energy stored in the magnetosphere of pulsar. We can have the energy loss rate per unit volume as~\cite{Korwar:2017dio}
\begin{eqnarray}
\frac{d^2 E_{\rm MCP}}{dtdV}=\Gamma_{\chi\overline{\chi}}^{\rm M}(\mathcal{E}(z))2m_\chi+ \Gamma_{\chi\overline{\chi}}^{\rm M}(\mathcal{E}(z))\int_z^h q_\chi \mathcal{E}(z') dz'\;,
\label{eq:loss}
\end{eqnarray}
where the energy losses due to MCPs are attributed to two sources. The first term accounts for the extracted energy per unit volume per unit time from the electric field for initiating MCP pair production.
We also define $E(z)\equiv \int_z^h q_\chi \mathcal{E}(z') dz'=q_\chi\mathcal{E}_{\rm max}(h-z)^2/(2h)$ as the work done by the Lorentz force on one MCP produced at height $z$ above the NS surface. This corresponds to the energy gained by the MCP while traversing the remaining distance $h-z$ under the electric field $\mathcal{E}(z)$. The second term thus gives the energy loss extracted from the electromagnetic field to accelerate MCPs to the top of the gap.
The magnetar's cumulative energy loss must not exceed its electromagnetic energy reservoir in the magnetar.
The mean energy loss rate of a magnetar over its lifetime $\mathcal{T}$ is thus constrained by
\begin{equation}
\int dV\Big[\frac{d^2E_{\rm rad}}{dt dV}+\frac{d^2E_{\rm MCP}}{dtdV}\Big]\lesssim
\int  dV\frac{1}{\mathcal{T}}\frac{\mathcal{B}^2}{2\mu_0}\;,
\label{eq:energyloss}
\end{equation}
where the left-hand side of Eq.~(\ref{eq:energyloss}) is the energy loss rate due to the magnetar radiation and the MCP production, the right-hand side is the magnetic energy stored in the magnetar with $\mu_0$ being the vacuum magnetic permeability~\footnote{Note that here we ignore the subdominant electric energy from perpendicular electric field in the magnetosphere.}, and $dV$ represents the differential volume element in the region under consideration. On the right-hand side, we consider the magnetic energy stored in magnetar interior and the magnetosphere of characteristic height $R_{\rm M}$. The dipole approximation~\cite{Abu-Ajamieh:2026spj}
\begin{eqnarray}
\mathcal{B}(r)=
\begin{cases}
\mathcal{B}\;,~~~r<R_{\rm M}\\
\mathcal{B}(R_{\rm M}/r)^3\;,~~~R_{\rm M}\leq r\leq 2R_{\rm M}
\end{cases}
\end{eqnarray}
with surface magnetic field $\mathcal{B}$ is used to model the magnetic field of magnetar.
We also use an active magnetar lifetime $\mathcal{T}\sim 10^4~{\rm yr}$ which corresponds to the magnetic field decay timescale~\cite{Colpi:1999gg,Beniamini:2019bga}. $\frac{d^2E_{\rm rad}}{dt dV}$ on the left-hand side of Eq.~(\ref{eq:energyloss}) represents the magnetar radiation loss rate per unit volume. For our calculation, we adopt the persistent quiescent X-ray luminosity of known magnetars in the McGill catalog~\cite{Olausen:2013bpa,McGillMagnetarCatalog} as shown in Table~\ref{tab:mcgill}
\begin{equation}
\int dV \frac{d^2E_{\rm rad}}{dt dV}=L_X\sim 10^{31}-10^{35}~{\rm ergs}~{\rm s}^{-1}\;.
\end{equation}
Recent observations continue to support the magnitude adopted for the persistent radiative losses of magnetars. The known Galactic magnetars exhibit persistent soft X-ray luminosities in the range of $L_X\sim 10^{31}-10^{36}~{\rm ergs}~{\rm s}^{-1}$~\cite{Shao:2024lnl,Ibrahim:2024egy}.
For typical magnetars with polar fields $\mathcal{B}_p\sim 4\times 10^{14}~{\rm G}$, the average crustal X-ray luminosity powered by magnetic field decay is estimated at $3\times 10^{34}~{\rm ergs}~{\rm s}^{-1}$~\cite{Soni:2018buv} which is consistent with the representative values employed in the present work.
Systematic re-analysis of the INTEGRAL/IBIS-ISGRI archive for more than two decades provides the most up-to-date characteristics of the persistent hard X-ray emission from the Galactic magnetar population~\cite{Pacholski:2024rbg}. It confirms that the quiescent X-ray luminosity outside burst-active periods remains in this range.

After subtracting the X-ray radiation from the magnetar energy reservoir and substituting the energy loss due to MCP Schwinger production in Eq.~(\ref{eq:loss}), we can obtain the constraints on the MCP electric charge fraction $\epsilon$ as a function of MCP mass $m_\chi$.

\section{Projected sensitivity and constraint}
\label{sec:results}

We first show the projected sensitivity of the laser-assisted Compton scattering to the MCP electric charge fraction. The input parameters are taken from the proposed Laser Und XFEL Experiment (LUXE)~\cite{Abramowicz:2019gvx,Abramowicz:2021zja} at the European X-Ray Free-Electron Laser (EuXFEL)~\cite{Altarelli:2006zza}. The electron beam is supposed to have the energy of $E_e=16.5$ GeV. It collides with
an intense laser beam of green light given $\omega=2.35$ eV~\cite{Bamber:1999zt,LUXE:2023crk} and the initial scattering angle $\theta=17^\circ$~\cite{Bamber:1999zt,LUXE:2023crk}. The signal event in our case is composed
of a single electron and missing energy carried away by the MCP pair, and becomes
\begin{eqnarray}
N_s = {1\over 2\rho_\omega} \Gamma_{\chi\overline{\chi}}^{\rm L} \cdot \mathcal{L}\;,
\end{eqnarray}
where $\rho_\omega={a^2 \omega\over 4\pi}$ is the laser photon density~\cite{Greiner:1992bv} with ${1\over 2\rho_\omega} \Gamma_{\chi\overline{\chi}}^{\rm L}$ being the scattering cross section~\cite{Greiner:1992bv}, and $\mathcal{L}$ is the integrated luminosity. The luminosity is given by
\begin{eqnarray}
\mathcal{L}=N_e \rho_\omega \ell N_b f t\;,
\end{eqnarray}
where $N_e=1.5\times 10^9$ denotes the number of electrons in a bunch in the electron beam of the EuXFEL accelerator for the LUXE experiment~\cite{LUXE:2023crk}, $N_b=2700$ is the number of individual bunches in the beam~\cite{LUXE:2023crk}, the laser operating frequency is $f=1~{\rm Hz}$~\cite{LUXE:2023crk}, the electron pathlength through
the laser focus is $\ell\simeq 50~{\rm \mu m}$~\cite{Bamber:1999zt}, and the physics data-taking time is $t=5\times 10^6~{\rm s}$ after considering the LUXE data taking efficiency of $75\%$~\cite{LUXE:2023crk}.
We obtain $\mathcal{L}\simeq 1.75~(700)~{\rm ab}^{-1}$ by choosing $\eta=0.5~(10)$ and the above parameters.
We also consider a higher energy benchmark with $E_e=125~{\rm GeV}$ as a representative scenario of future high-energy electron beam~\cite{Irles:2023udu,Schulthess:2025tct}. The integrated luminosity is assumed to be the same as that for LUXE.
Given the SM background calculated above, we require a
$3\sigma$ observation based on the significance formula~\cite{ParticleDataGroup:2026aaa}
\begin{eqnarray}
{N_s\over \sqrt{N_s+N_b}}\;,
\end{eqnarray}
where $N_s$ ($N_b$) denotes the number of signal (background) events. The background event count $N_b$ is obtained by substituting $\Gamma_{\chi\overline{\chi}}^{\rm L}$ with $\Gamma_{\nu\overline{\nu}}^{\rm L}$ in $N_s$.

\begin{figure}[htbp]
\centering
\includegraphics[width=0.6\textwidth]{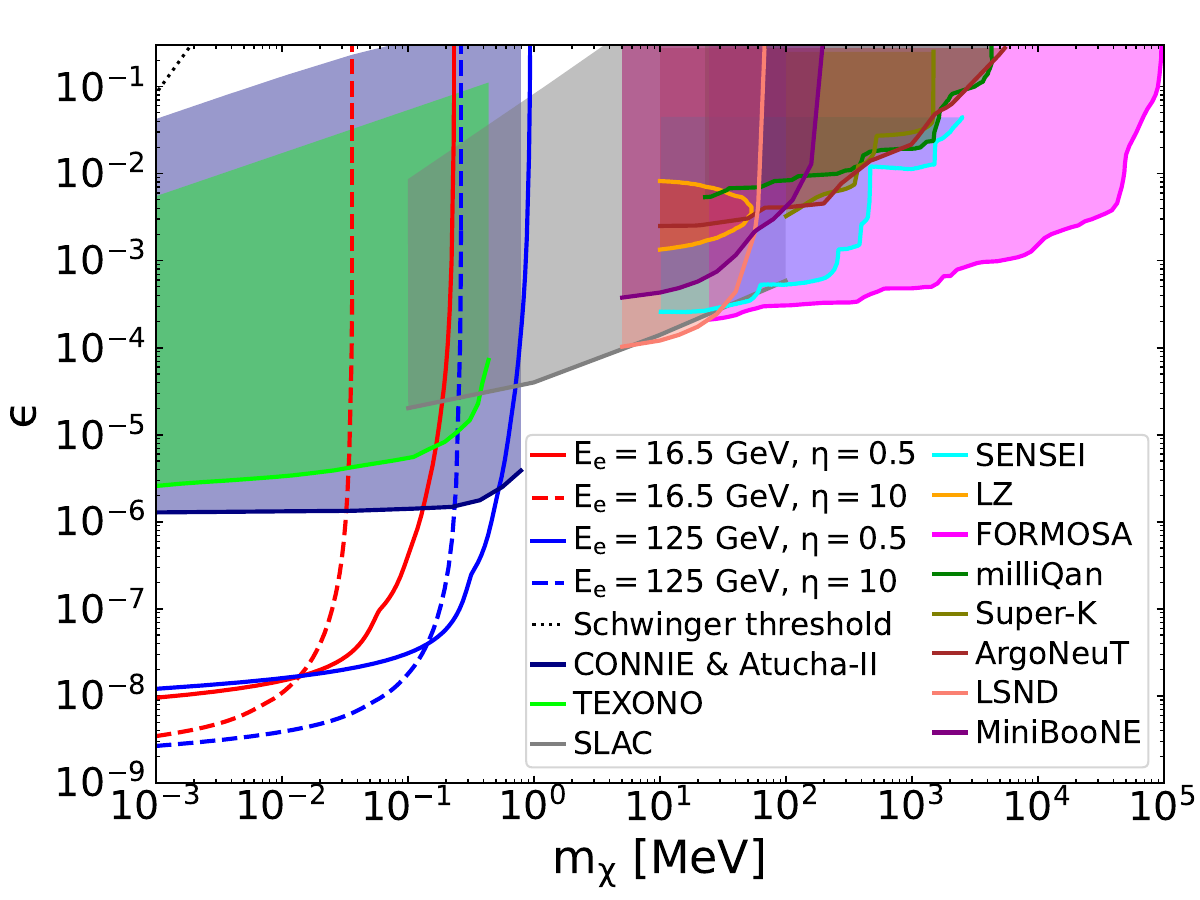}
\caption{Sensitivity of nonlinear Compton scattering to the MCP electric charge fraction $\epsilon$ as a function of $m_\chi$. We take $E_e=16.5$ GeV (red lines) or $E_e=125$ GeV (blue lines) and set $\eta=0.5$ (solid lines) or $\eta=10$ (dashed lines). The exclusion regions from other experiments are also
shown for comparison, including SLAC (gray region)~\cite{Prinz:1998ua}, LSND (salmon region)~\cite{Magill:2018tbb}, MiniBooNE (purple region)~\cite{Magill:2018tbb}, TEXONO (lime region)~\cite{TEXONO:2018nir}, ArgoNeuT (brown region)~\cite{ArgoNeuT:2019ckq}, SuperK (olive region)~\cite{Plestid:2020kdm}, milliQan (green region)~\cite{Ball:2020dnx}, SENSEI (cyan region)~\cite{SENSEI:2023gie}, combined CONNIE and Atucha-II (navy region)~\cite{CONNIE:2024off}, LZ (orange region)~\cite{LZ:2025xkj}, and FORMOSA (magenta region)~\cite{Citron:2025kcy}. The dotted line yields the MCP critical field strength $\mathcal{E}= \mathcal{E}_{\rm cri.}=m_\chi^2/(\epsilon e)$ for $\eta=10$.
}
\label{fig:laserlimit}
\end{figure}

Fig.~\ref{fig:laserlimit} shows the sensitivity of nonlinear Compton scattering to the MCP electric charge fraction $\epsilon$ as a function of $m_\chi$. We take $E_e=16.5$ GeV (red lines) or $E_e=125$ GeV (blue lines) and set $\eta=0.5$ (solid lines) or $\eta=10$ (dashed lines) for illustration. The exclusion regions from other experiments are also
shown for comparison~\footnote{Although the literature figures indicate a lower mass cutoff, the actual constraint data may be valid at even lower masses.}. The dotted line marks where the field strength of green laser with $\eta=10$ reaches the MCP critical threshold $\mathcal{E}= \mathcal{E}_{\rm cri.}=m_\chi^2/ q_\chi$ with $q_\chi=\epsilon e$~\footnote{The line for $\eta=0.5$ is above the one for $\eta=10$ and is not shown in this figure.}. The MCPs with parameters above this dotted line would be spontaneously produced in pairs through the Schwinger effect. One can see that the projected limit can reach as low as $\epsilon\sim  10^{-8}~(3\times 10^{-9})$
for $\eta=0.5~(10)$. As $\eta$ increases from 0.5 to 10, both the strong-field correction to the effective MCP mass $m_\chi^\ast$ and the laser-assisted decay width $\Gamma_{\chi\overline{\chi}}^{\rm L}$ increase. As a result, the maximum reachable MCP mass $m_\chi$ and the corresponding upper limit on $\epsilon$ decrease. The laser-assisted
process provides a complementary search of MCP for $m_\chi \lesssim 1~{\rm MeV}$, compared with other
laboratory experiments.

\begin{figure}[htbp]
\centering
\includegraphics[width=0.7\linewidth]{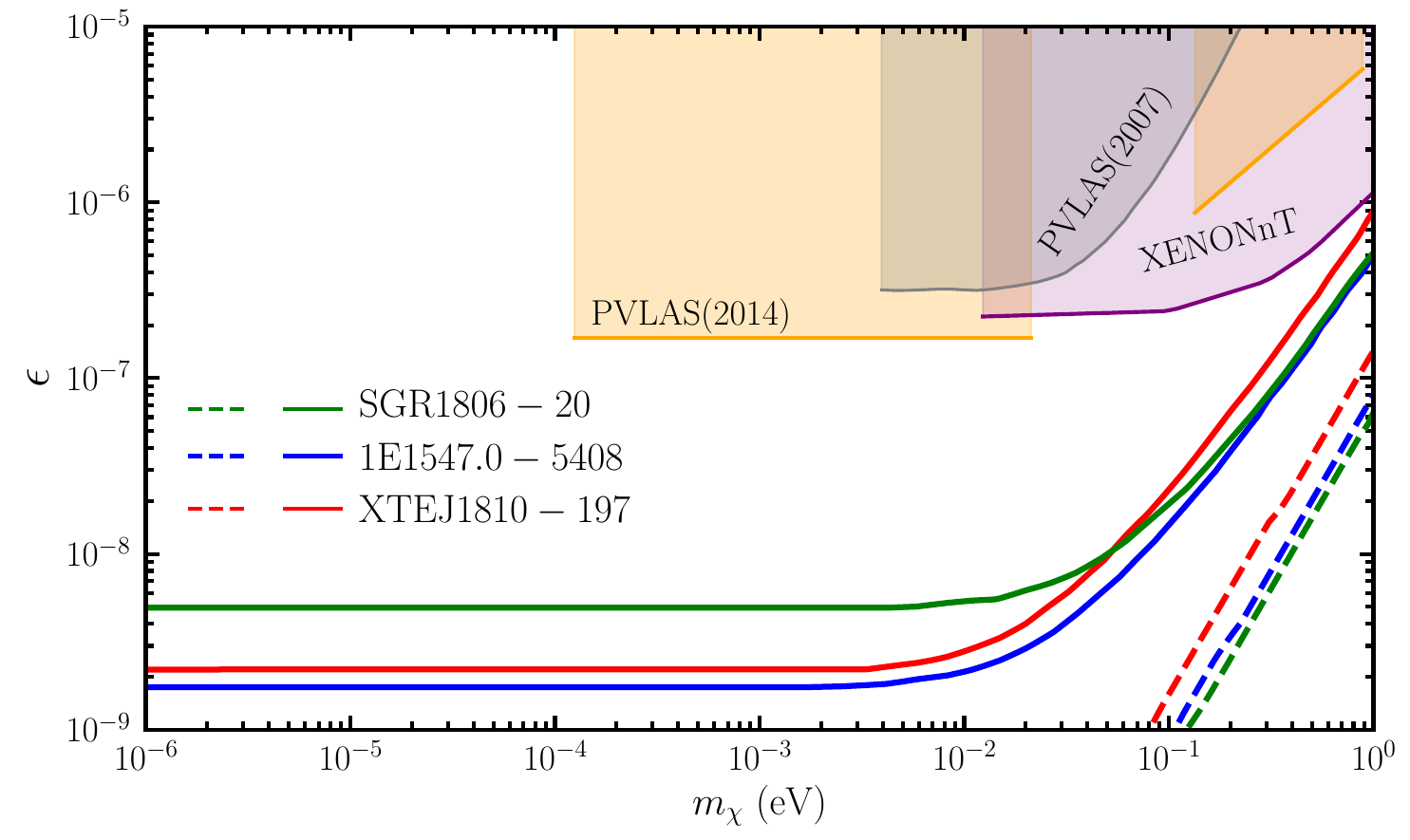}
\caption{Constraints on the MCP electric charge fraction $\epsilon$ from energy loss induced by Schwinger production in magnetars XTE J1810-197 (red solid line), 1E 1547.0-5408 (blue solid line) and SGR 1806-20 (green solid line). The dashed line of the same color indicates the typical threshold of MCP Schwinger pair production for the corresponding magnetar, satisfying $\Gamma_{\chi\overline{\chi}}^{\rm M}=1~{\rm m}^{-3}~{\rm s}^{-1}$ with $\mathcal{E}=\mathcal{E}_{\rm max}$. We also show the exclusion regions from PVLAS (2007, gray region)~\cite{Ahlers:2007qf}, PVLAS (2014, orange region)~\cite{DellaValle:2014xoa} and XENONnT (purple region)~\cite{Kouvaris:2025tom} for comparison.
}
\label{fig:astroexclusion}
\end{figure}

In Fig.~\ref{fig:astroexclusion}, we show the constraints on the MCP electric charge fraction $\epsilon$ from energy loss induced by Schwinger production in magnetars XTE J1810-197 (red line), SGR 1806-20 (green line) and 1E 1547.0-5408 (blue line) for illustration. One can see that the upper limit of MCP electric charge fraction can reach as low as $\epsilon\sim 10^{-9}$ for $m_\chi\lesssim 10^{-2}~{\rm eV}$. This is weaker than the bound in Ref.~\cite{Korwar:2017dio} by using approximate expressions for polar gap properties. The dashed line of the same color yields $\Gamma_{\chi\overline{\chi}}^{\rm M}=1~{\rm m}^{-3}~{\rm s}^{-1}$ with $\mathcal{E}=\mathcal{E}_{\rm max}$ and indicates the typical threshold of Schwinger pair production for the corresponding magnetar. Below each dashed line, the Schwinger pair production of MCP is dramatically suppressed by $\mathrm{exp}(-\frac{\pi m^2_\chi}{q_\chi\mathcal{E}})$. The existing exclusion regions from PVLAS (2007, gray region)~\cite{Ahlers:2007qf}, PVLAS (2014, orange region)~\cite{DellaValle:2014xoa} and XENONnT (purple region)~\cite{Kouvaris:2025tom} are also shown for comparison. The bounds for light MCPs from energy loss induced by Schwinger production in magnetars are two orders of magnitude stronger than a terrestrial apparatus such as PVLAS or XENONnT.

\section{Conclusion}
\label{sec:Con}

The search for light dark particles under terrestrial and astrophysical strong-field environments has drawn much attention. High-intensity laser pulses and highly magnetized NSs provide ideal strong-field environments for detecting dark sector candidates.
In this work, we investigate the potential to search for and constrain light MCPs via strong
electromagnetic fields in both laboratory laser experiments and astrophysical magnetars.

We first suggest the MCP pair production from
nonlinear Compton scattering through the interaction of an ultra-relativistic electron beam with a high-intensity laser pulse. The Furry picture
and Volkov solution of Dirac equation in an external electromagnetic background field are used to describe the electrons and MCPs in a classical laser field. We calculate the cross sections of nonlinear
Compton scattering to MCPs and obtain the sensitivity reach of MCP electric charge fraction by taking into account the irreducible SM background with missing neutrinos. We also reexamine the Schwinger pair production of MCPs from magnetars with ultra-strong magnetic field and parallel electric field in polar gap. The properties of polar gap are outlined based on Ruderman-Sutherland model for confirmed magnetars in McGill catalog. We evaluate the energy loss due to the MCP pair production and the consequent acceleration by strong electric field.

We find the follow conclusions
\begin{itemize}
\item The projected limit of MCP electric charge fraction $\epsilon$ from nonlinear Compton scattering can reach as low as $\epsilon\sim  10^{-8}~(3\times 10^{-9})$
with intensity parameter $\eta=0.5~(10)$. Compared with other
laboratory experiments, the laser-assisted nonlinear
process provides a complementary search of MCP for $m_\chi \lesssim 1~{\rm MeV}$.
\item The upper limits of MCP electric charge fraction from energy loss induced by Schwinger production in magnetars can reach $\epsilon\sim 10^{-9}$ for $m_\chi\lesssim 10^{-2}~{\rm eV}$.
The bounds are two orders of magnitude stronger than a terrestrial apparatus such as PVLAS or XENONnT.
\item Laboratory laser experiments are advantageous for probing MCP parameter regions below the critical field strength, whereas the Schwinger production in magnetars is more favorable for exploring regions above it. The constraints from highly magnetized magnetars and the search
potential in laser experiments are complementary.
\end{itemize}

\acknowledgments

We would like to thank Chen Sun for useful discussions.
T.~L. is supported by the National Natural Science Foundation of China (Grant No. 12375096).
K. M. is supported by the Shaanxi Fundamental
Science Research Project for Mathematics and Physics (Grant No. 25JSY031)


\bibliography{refs}

\end{document}